# Performance of Dual-Augmented Lagrangian Method and Common Spatial Patterns applied in classification of Motor-Imagery BCI

A. Miladinović[1], M. Ajčević[1], G. Silveri[1], A. Accardo[1]
[1] Department of Engineering and Architecture, University of Trieste, Italy

*Abstract*—Motor-imagery based brain-computer interfaces (MI-BCI) have the potential to become ground-breaking technologies for neurorehabilitation, the reestablishment of non-muscular communication and commands for patients suffering from neuronal disorders and disabilities, but also outside of clinical practice, for video game control and other entertainment purposes. However, due to the noisy nature of the used EEG signal, reliable BCI systems require specialized procedures for features optimization and extraction. This paper compares the two approaches, the Common Spatial Patterns with Linear Discriminant Analysis classifier (CSP-LDA), widely used in BCI for extracting features in Motor Imagery (MI) tasks, and the Dual-Augmented Lagrangian (DAL) framework with three different regularization methods: group sparsity with row groups (DAL-GLR), dual-spectrum (DAL-DS) and l1-norm regularization (DAL-L1). The test has been performed on 7 healthy subjects performing 5 BCI-MI sessions each. The preliminary results show that DAL-GLR method outperforms standard CSP-LDA, presenting 6.9% lower misclassification error (p-value = 0.008) and demonstrate the advantage of DAL framework for MI-BCI.

## I. Introduction

THE purpose of BCI research is to establish communication between computers and the brain bypassing peripheral nerves. The approach can be used to regain communication capabilities in the cases of certain types of neurological conditions [1], control of a prosthetic arm or wheelchair [2], and furthermore, by combining different strategies [3] as neurorehabilitation tools for patients suffering from Parkinson's disease [4], post-Stroke [5], Attention-Deficit Hyperactivity Disorders [6], Autism Spectrum Disorder [7] and other neurological conditions.

The most commonly used BCI technique is based on electroencephalography since it provides a non-invasive, easily applicable, and relatively affordable measurement of the electrical activity generated on the scalp as a consequence of the brain activity. However, despite its advantages, the acquired signal quality may vary due to the low signal-to-noise ratio caused by external factors such as powerline of 50Hz and man-made noise, but also internal non-task related brain and muscular activity. Nonetheless, the signal has high time-resolution; it suffers from low spatial resolution caused by tissue volume conduction and non-stationarity caused by changes in the subject's overall brain state during the experiment (e.g., a prominent increase of alpha burst caused by drowsiness). Therefore, the application of conventional statistical analysis leads to poor performance and, in many cases, fail to produce a reliable real-time prediction of the user's brain state. Therefore, one of the biggest challenges in BCI is how to extract a small number of robust and informative features from the EEG data by applying various data-driven approaches. One of the most commonly used techniques for the transformation of sensor space (raw EEG data) into a new lower rank space is Common Spatial Patterns (CSP) [8]. Various studies [8]–[10] demonstrated that the pre-processing steps mostly influence the performance of the BCI system, and in that regard, it is demonstrated in [11] that relatively simple classifiers, such as Linear Discriminant Analysis (LDA) [8], can provide optimal results and outperform more complex classification techniques. Still, most data-driven approaches, including CSP, as reported in [12] tries to solve non-convex optimisation problems, and subdivided independent steps of feature extraction, and classification increases the chances of suboptimal BCI models leading to general poorer performances and lower resilience to the noises [13].

In that regard, in this paper we explored the performance of an alternative framework, the Dual-Augmented Lagrangian (DAL) [14], [15], that tries to eliminate problems of multiple local minima by merging the process of feature extraction, dimensionality reduction, and classification into one process [14]. The further aim is to provide a comparison of the performance of CSP and DAL for motor-imagery based brain-computer interface in a real-life scenario applied to a group of healthy subjects.

## II. Materials and Methods

*A. Study population*

The experiment was conducted on 7 healthy (3 males and 4 females) with a mean age of 21 years (standard deviation= 1.6) right-handed (Edinburgh Handedness Inventory) participants.

All subjects were right-handed, with no history of neurological disorders and BCI naïve.

This study has been approved by the local ethics committee and has been performed in accordance with the ethical standards laid down in the 1964 Declaration of Helsinki and its later amendments. All participants released their informed consent.

*B. BCI protocol*

Each subject underwent 5 sessions of right-hand Motor-Imagery controlled BCI feedback administered in two weeks.

Each session consisted of initial calibration phase in which subjects had a task to perform a repetitive MI right-hand task when the command in the form of the arrow appeared on the screen in the duration (see Fig.1), and, conversely, stay in rest when the image disappears (two-class classification problem). The number of repetitions was set to 30 for each class, leading to a total of 60 acquired trials for each session. The acquired data is then used for the subsequent online phase where the cursor was controlled by the subject's volitional EEG processed with the produced BCI model.

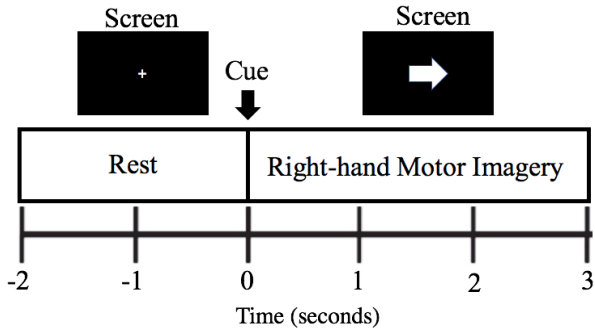

Fig. 1: Depiction of the presented stimulus on the screen for a single run during initial calibration phase

*C. EEG acquisition*

The acquisition of 11 channel EEG was performed using OpenBCI (OpenBCI, New York, NY, USA) and Ag/AgCl electrodes (SpesMedica, Genova, Italy) placed at standard 10-20 locations covering the motor areas (F3, Fz, F4, T3, C3, Cz, C4, T4, P3, Pz, P4) as reported in Fig.2. The signals were recorded with 250 Hz sample frequency and subsequently downsampled to 128Hz and pre-processed with the 6-32 Hz 2nd order Butterworth digital bandpass filter. During the experiment, the electrode impedances were kept below 5 kΩ. The BCILAB [16] and MATLAB® (The MathWorks Inc., Natick, MA) were used for real-time processing, as well as for later offline analysis.

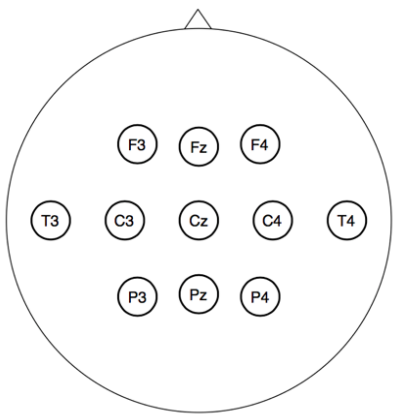

Fig. 2: Schematic of electrode placement for MI-BCI

The performance of DAL has been explored with tree type of regularisation strategies proposed in [14] applicable to the augmented weight matrix: (1) group sparsity with row groups (DAL-GLR), suggested in the cases where the emphasis is on channel selection, and there are a low number of temporally distributed activity patterns, (2) dual-spectral optimal where there are few informative constellations over time combining both power and time-domain segments (DAL-DS), and finally, (3) standard Euclidean l1-norm regularization that focuses on the temporal aspect.

At the same time, the CSP-LDA algorithm has been used as described in previous work [9], maximizing signal variance (power) for 3 patterns for each class. After the application of the spatial filter, the data are fed in the LDA classifier with automatic shrinkage regularization parameters [8].

The optimization of regularization parameters in all cases was performed using 5-fold blockwise cross-validation with 5 trials, whereas the final misclassification error of the model was estimated using 10-fold blockwise cross-validation with the same margin of 5 trials.

*D. Statistical analysis*

Differences in misclassification errors were tested by repeated-measure analysis of variance (ANOVA). Bonferroni corrections were used for post-hoc multiple comparisons.

### III. RESULTS

The session mean misclassification error obtained by models produced by CSP-LDA, DAL-GLR, DAL-DS, and DAL-L1 approaches is reported in Table 1 for each of the 7 subjects. The misclassification error of the DAL-GLR (30.7% ± 6.3) was significantly lower compared to CSP (37.6 ± 8.7) (p-value = 0.008). No significant difference in total error was found between CSP-LDA and DAL-DS (p-value = 0.060), and DAL-L1 (p-value = 0.095), although, on average, both cases presented a slightly lower error. No significant differences were observed among DAL-based methods

### IV. DISCUSSION

Motor-Imagery BCI has the potential to be used as a neurorehabilitation tool that can improve motor and cognitive abilities in patients and subjects affected by various disorders [4]–[7].

TABLE I
MISCLASSIFICATION ERROR (%) AND STANDARD DEVIATION OBTAINED BY THE CSP-LDA, DAL-GLR, DAL-DS, AND DAL-L1 METHODS, RESPECTIVELY, FOR 7 HEALTHY SUBJECTS OVER 5 BCI SESSIONS PERFORMED FOR EACH SUBJECT. FOR EACH PATIENT, THE LOWEST ERROR IS MARKED IN BOLDFACE.

| Subjects | CSP-LDA | DAL-GLR | DAL-DS | DAL-L1 |
|---|---|---|---|---|
| 1 | 40.1±4.9 | 25.2±0.3 | **24.9**±0.3 | 25.3±0.7 |
| 2 | 47.6±3.5 | 25.7±1.0 | **24.9**±0.4 | 25.4±1.3 |
| 3 | 37.9±5.1 | **32.5**±5.7 | 35.8±6.4 | 33.4±5.5 |
| 4 | 37.0±9.5 | **33.9**±11.4 | 34.3±8.2 | 34.4±12.0 |
| 5 | **25.1**±4.9 | 30.8±3.9 | 30.8±7.0 | 33.6±8.2 |
| 6 | **31.5**±4.8 | 31.5±5.5 | 33.7±7.8 | 31.9±4.5 |
| 7 | 44.3±4.0 | **35.2**±4.9 | 38.4±6.2 | 38.0±5.7 |
| **Average** | 37.6±8.7 | **30.7**±6.3 | 31.8±7.4 | 31.7±7.4 |

To improve the reliability of BCI systems, a proper model that personalizes the model for each subject, and each session has to produce. The violation of the Machine Learning premises, such as unpredicted non-stationarity of the features [13], imposes a challenge to produce a reliable BCI model using a standard optimization technique. Therefore, the developed frameworks have to be also properly tested and evaluated on subjects producing real-life application scenarios.

This study compared the performance of two approaches for two-class MI-based BCI, the traditional CSP algorithm, and the DAL method with three different regularisation strategies. The study was performed on 7 healthy subjects. The approach DAL with the group sparsity regularization DAL-GLR outperforms standard CSP-LDA algorithm when the task elicits changes in power band ratios at the particular scalp locations, such as mu/alpha power decrease and beta increase (beta rebound) in the case of Motor-Imagery [2]. The execution of Motor Imagery activates Sensory-motor areas that are located contralaterally of the imagined hand [10], and in our case, the highest activation produces by right-hand is expected on the C3 electrode. The precise topographical location of expected cortex activation explains why DAL-GLR, optimized for channel selection, produces the lowest misclassification error on average.

The better accuracy DAL in respect to the CSP of 6.9% (p-value = 0.008) demonstrate the superiority of the framework that blurs the boundary between subsequent machine-learning steps because it provides an immediate loop for optimization of regularization parameters, which is not the case in CSP where the feature extraction and reduction are separated from the classification process.

Nonetheless, due to a small test, it is difficult to draw any firm conclusions for the other two regularization approaches DAL-DS and DAL-L1, despite statistically nonsignificant differences also present lower error in comparison to CSP.
Evidence from [14] shows DAL-DS outperforms the DAL-GLR when applied to slow cortical potentials. However, in the case of our experiment, the findings demonstrate activation of the precise brain location and on higher frequencies (alpha/beta), which is not demonstrated in the case of the study in [14]. Despite not optimal in all cases, we can observe that the DAL-DS presented the lowest classification error in two of seven subjects, which can lead to a debate that the differences among DAL approaches should be further investigated in a larger dataset.

## V. Conclusion

In conclusion, this preliminary study showed that among selected approaches, DAL-GLR provides the best classification performance among the tested approach. It also demonstrates the supremacy of DAL methods over the standardized CSP-LDA machine learning framework. However, the results obtained should be confirmed in a study with a larger number of participants

### Acknowledgement

A. Miladinović is supported by the European Social Fund (ESF) - FVG.

Work partially supported by master's programme in Clinical Engineering of the University of Trieste.